\documentclass[12pt, preprint]{aastex}

\usepackage{graphicx}
\usepackage{amssymb}
\usepackage{amsmath}
\usepackage{layout}
\usepackage{url}
\usepackage{subfigure}
\usepackage{color}
\usepackage{enumerate}
\newcommand{\vect}[1]{\boldsymbol{#1}}

\slugcomment{Submitted to {\it The Astrophysical Journal Letters}}

\begin{document}


\title{Large-Amplitude Blazar Polarization Angle Swing as a Signature of Magnetic Reconnection}


\author{Haocheng Zhang\altaffilmark{1}, Xiaocan Li\altaffilmark{2}, Fan Guo\altaffilmark{2}, Dimitrios Giannios\altaffilmark{1}}

\altaffiltext{1}{Department of Physics and Astronomy, Purdue University, West Lafayette, IN 47907, USA}

\altaffiltext{2}{Theoretical Division, Los Alamos National Laboratory, Los Alamos, NM 87545, USA}

\begin{abstract}

Relativistic magnetic reconnection events can widely exist in magnetized plasmas in astrophysical systems. During this process, oppositely directed magnetic field lines reconnect and release magnetic energy, efficiently accelerating nonthermal particles. However, so far there is little clear observational signatures of relativistic magnetic reconnection events in astrophysical systems. Blazars are relativistic magnetized plasma outflows from supermassive black holes. Their multi-wavelength flares may be powered by relativistic magnetic reconnection. The highly variable radiation and polarization signatures are well covered by multi-wavelength observation campaigns, making them ideal targets to examine the magnetic reconnection model. Recent observations have found that several blazar flares are accompanied by optical polarization angle swings which may be of as large amplitude as $> 180^{\circ}$, challenging existing theoretical models. In this paper, we present integrated particle-in-cell (PIC) and polarized radiation transfer simulations of magnetic reconnection events. We find that plasmoid coalescences in the reconnection layer can give rise to highly variable light curves, low and fluctuating polarization degree, and rotating polarization angle. In particular, large-amplitude polarization angle swings, similar to those observed during blazar flares, can be a unique signature of relativistic magnetic reconnection events.
\end{abstract}


\section{Introduction}

Magnetic reconnection is a plasma physics process ubiquitously occurring in space and astrophysical environments where oppositely directed field lines break  and rejoin. During reconnection a large amount of magnetic energy can be released, especially under strong magnetic field conditions. This is particularly important for magnetically dominated astrophysical systems. Recent simulations have shown that relativistic magnetic reconnection can result in efficient particle acceleration \citep{Sironi14,Guo14,Guo15,Werner16}. 

Magnetic reconnection has been extensively studied in the nonrelativistic regime, where in situ measurements and solar flare imaging observations have provided much detail \citep{Phan00,Gosling05,Tian14,Wang16}. Magnetic reconnection may also widely exist in high-energy astrophysical systems such as pulsar wind nebulae and relativistic jets from black holes. Blazar jets are relativistic plasma outflows that are launched from the central supermassive black hole with considerable magnetic energy \citep{Blandford77}. During the jet propagation, magnetohydrodynamical instabilities may trigger magnetic reconnection which dissipates magnetic energy accelerating nonthermal particles \citep[see, e.g.,][]{Giannios09}. Observationally, blazars have shown very fast $\gamma$-ray variability \citep[e.g.,][]{Ackermann16} and very hard photon spectra, which are promising evidence of magnetic reconnection \citep{Petropoulou16}. However, unlike reconnection in Earth's magnetosphere, one cannot directly measure the magnetic field in blazar jets, thus so far there has been little conclusive  evidence of relativistic magnetic reconnection events in blazar jets.

Polarimetry is a standard probe of the astrophysical magnetic field. Since the optical blazar emission is dominated by synchrotron of nonthermal electrons, polarization is then a direct measurement of the apparent magnetic field in the emission region. Blazar optical emission exhibits strong variability in both flux and polarization signatures. In particular, recent optical polarization monitoring programs have discovered optical polarization angle (PA) swings that are frequently accompanied by multi-wavelength flares \citep{Marscher08,Marscher10,Blinov16,Blinov18}. Theoretical models usually suggest $\lesssim 180^{\circ}$ PA swings that originate from physical processes altering the partially ordered magnetic field in the emission region \citep[e.g.,][]{Marscher08,Marscher14,ZHC16}. However, an extreme kind of events, multi-wavelength flares with simultaneous large-amplitude ($> 180^{\circ}$) PA swings, are more challenging to account for. Similar to their $\lesssim 180^{\circ}$ counterparts, these events show one or multiple flares as well as low and fluctuating polarization degree (PD). But here the PA can rotate much more than $180^{\circ}$, either consistently in one direction \citep{Marscher10} or in both directions \citep{Chandra15}. Because of the large-amplitude and rather smooth PA rotation, they are unlikely due to stochastic processes in a turbulent magnetic field. Instead, these features indicate a highly dynamical but regulated alteration of the magnetic field morphology.

In this paper, we present a study of radiation and polarization signatures from relativistic magnetic reconnection by an integrated modeling relying on first-principle particle-in-cell kinetic simulation and an advanced polarized radiation transfer simulation. Our goal is to establish the physical link between magnetic reconnection and large-amplitude PA swings in blazars, and understand what dynamical features in the reconnection layer produce the PA swing. Section \ref{model} describes our simulation setup, section \ref{result} presents the results, and section \ref{discussion} discusses our findings and summarizes the paper.

\section{Model Description \label{model}}

We perform 2D PIC simulations in the $x-z$ plane using the VPIC code \citep{Bowers08}. We employ periodic boundary conditions in the $x$-axis for both fields and particles, while in the $z$-axis the boundaries are conductive for fields but reflect particles. The simulation starts from a magnetically-dominated force-free current sheet, $\textbf{B}=B_0\text{tanh}(z/\lambda)\hat{x}+B_0\text{sech}(z/\lambda)\hat{y}$. This corresponds to a rotating magnetic field with a $180^\circ$ change in direction within a thickness of $2 \lambda$. We set the half-thickness $\lambda$ of the current sheet to be 120 $d_{e0}$, where $d_{e0}=c/\omega_{pe0}$ is the nonrelativistic electron inertial length and $\omega_{pe0}=\sqrt{4\pi n_ee^2/m_e}$ is the nonrelativistic electron plasma frequency, so that the electron motion can support the current density. The initial particle distributions are spatially uniform with relativistic Maxwellian in energy space. The simulation assumes an electron-ion plasma with realistic mass ratio $m_i/m_e=1836$. We use $100$ electron-ion pairs in each cell. We insert a long-wavelength perturbation to trigger the magnetic reconnection, which creates a dominating reconnection point located at the center of the simulation box \citep{Birn01}.

Observationally, flat-spectrum radio quasars (FSRQs) generally exhibit the strongest variability and polarized variability \citep{Ackermann16,Angelakis16}. Here we try to mimic the physical conditions of a typical FSRQ emission region. Fits to blazar spectra suggest that the low-energy cutoff of the nonthermal electron Lorentz factor distribution ranges from hundred to thousand \citep{Boettcher13}, which may correspond to their thermal temperature. For simplicity, the initial thermal temperatures for ions and electrons are assumed to be $T_i=T_e=100~m_ec^2$. FSRQs usually have very strong cooling due to the synchrotron and Compton scattering. Here we mimic the cooling effect by implementing a radiation reaction force $\vect{g}$ in VPIC, which can be simplified as a continuous friction force for ultra-relativistic particles \citep{Cerutti12, Cerutti13}.
\begin{align}
  \vect{g} & = -\frac{\mathcal{P}_\text{rad}}{c^2}\vect{v}
  = -\frac{2}{3}r_e^2\gamma\left[\left(\vect{E}+
  \frac{\vect{u}\times\vect{B}}{\gamma}\right)^2 -
  \left(\frac{\vect{u}\cdot\vect{E}}{\gamma}\right)^2\right]\vect{u},
\end{align}
where $\vect{u}$ is the four-velocity, $\mathcal{P}_\text{rad}$ is the radiation power radiated by a particle in an electromagnetic field and $r_e=e^2/m_ec^2$ is the classical radius of the electron. We normalize the equation of motion as
\begin{align}
  \tilde{m}\frac{d\vect{u}}{d\tilde{t}} & = \tilde{q}\left(\tilde{\vect{E}}+
  \frac{\vect{u}\times\tilde{\vect{B}}}{\gamma}\right)
  \frac{t_0eB_0}{m_ec} -\frac{2}{3}\tilde{r}_e^2\gamma\left[\left(\tilde{\vect{E}}+
    \frac{\vect{u}\times\tilde{\vect{B}}}{\gamma}\right)^2 -
  \left(\frac{\vect{u}\cdot\tilde{\vect{E}}}{\gamma}\right)^2\right]
  \frac{c^2t_0^3B_0}{m_ec}\vect{u} \nonumber \\
  & = \tilde{q}\left(\tilde{\vect{E}}+
  \frac{\vect{u}\times\tilde{\vect{B}}}{\gamma}\right)
  t_0\Omega_{ce0} -\frac{2}{3}\gamma\left[\left(\tilde{\vect{E}}+
    \frac{\vect{u}\times\tilde{\vect{B}}}{\gamma}\right)^2 -
  \left(\frac{\vect{u}\cdot\tilde{\vect{E}}}{\gamma}\right)^2\right]
  t_0^2\Omega_{ce0}^2\tilde{r}_e\vect{u}
\end{align}
where $\tilde{m}=m/m_e$, $\tilde{q}=q/e$, $\tilde{t}=t/t_0$, $\tilde{\vect{B}}=\vect{B}/B_0$, $\tilde{\vect{E}}=\vect{E}/B_0$, $\tilde{r}_e=r_e/(ct_0)$, $t_0=\omega_{pe0}^{-1}$, and $\Omega_{ce0}=eB/(m_ec)$ is the nonrelativistic electron gyrofrequency. Spectral fitting suggests that the FSRQ magnetic energy can be relatively strong and its high-energy nonthermal electron cutoff is $\gamma\sim 10^4$ \citep{Boettcher13}. Since the high-energy electron cutoff is roughly equal to the electron magnetization factor $\sigma_e\equiv B^2/(4\pi n_em_ec^2)=(\Omega_{ce0}/\omega_{pe0})^2$ \citep{Sironi14,Guo14}, we choose the total magnetization $\sigma_0\sim(m_e/m_i)\sigma_e\sim 22$, then $\sigma_e\sim 4\times 10^4$. The particle cooling time scale is given by $\tau_\text{cool} = 3t_0/(2\gamma\sigma_e\tilde{r}_e)$, which is set to be $1000~t_0$. The simulation box size is $2L\times L$ in the $x-z$ plane, where $L=8000d_{e0}$. Typical magnetic field strength in the FSRQs is $\sim 0.1~\rm{G}$. Thus our box size is normalized to $\sim 3\times 10^{10}~\rm{cm}$. While this is much smaller than the typical blazar emission region ($\sim 10^{16}$ cm), we find that the general plasmoid dynamics are qualitatively the same with domain size $2\times$ larger and smaller than the present case. Since the key mechanism in producing radiation signatures is the plasmoid coalescence (details in Section \ref{result}), this suggests the underlying process is robust even on the macroscopic scale. We choose a simulation grid size of $4212\times2106$, so that the cell sizes $\Delta x=\Delta z\sim 0.31d_e$ can resolve the thermal electron inertial length $d_e=\sqrt{\gamma_0}d_{e0}$, where $\gamma_0=1+3T_e/2m_ec^2\sim 150$.

The light crossing time scale in the $z$-axis of the simulation box is $\tau_{lc}=8000~t_0$. We output the simulation data every $125t_0\sim 0.016~\tau_{lc}$ to study the time-dependent radiation signatures. We record the particle energy spectra and averaged magnetic field in every zone domain with $\delta x \times \delta z \sim 150 d_{e0} \times 150 d_{e0}$, which is small enough to capture the main dynamic features in the simulation. The time-dependent energy spectra and magnetic field serve as inputs for the polarized radiation transfer code 3DPol.

The 3DPol code is a polarized radiation transfer code \citep{ZHC15}. The code evaluates the Stokes parameters from each of emission zones based on the magnetic field and nonthermal particle distribution, then traces the Stokes parameters to the plane of sky. Thus all light crossing time effects are naturally included. Recent upgrades include time-dependent polarized emission map, which can be used to pinpoint the connection between dynamics in the reconnection layer and polarized radiation signatures. By adding up the emission that arrives to the same plane of sky cell at the same time, it derives the spatially resolved polarized emission map at each time step.

\section{Results \label{result}}

\begin{figure}[ht]
\centering
\includegraphics[width=0.5\linewidth]{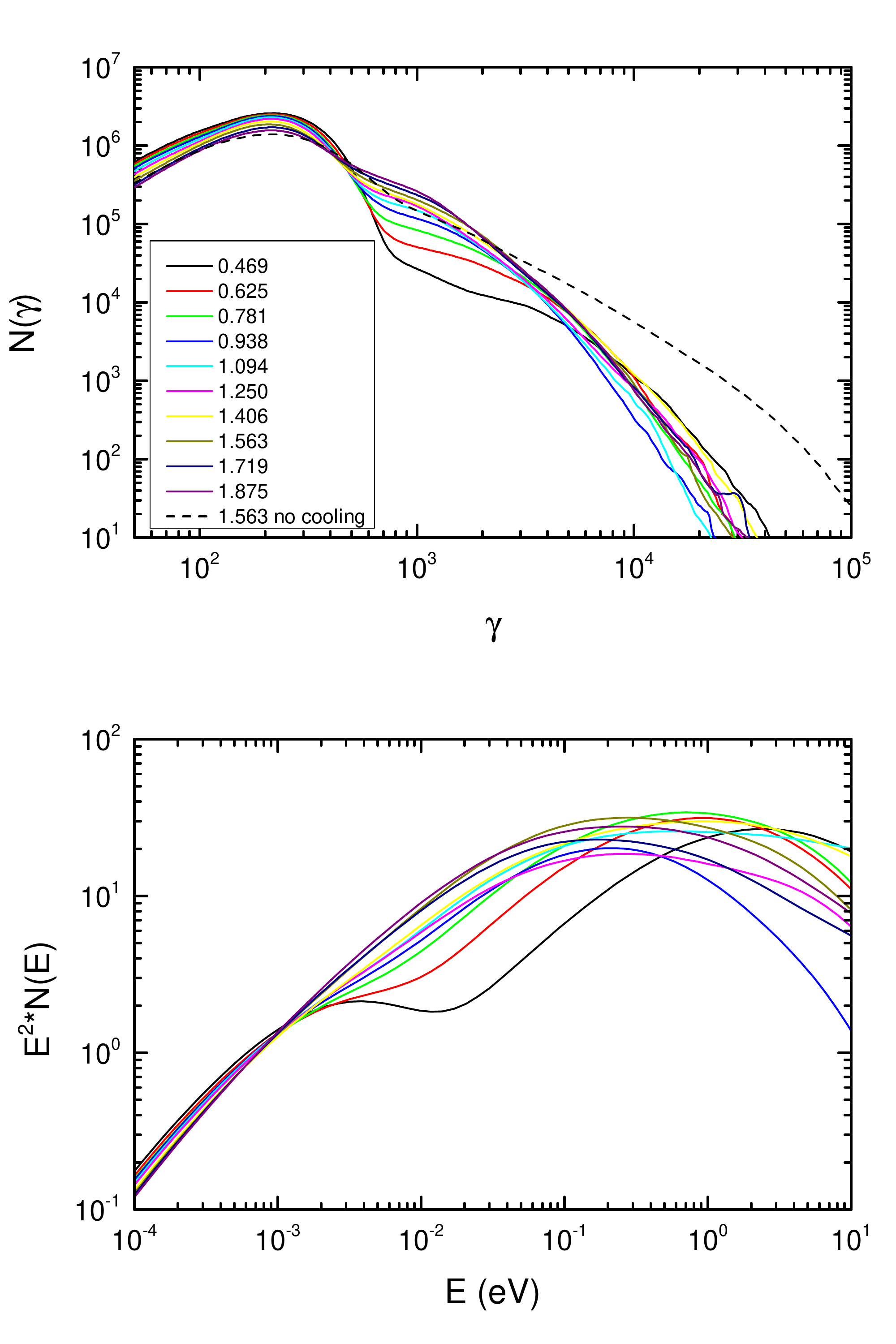}
\caption{Upper panel: time-dependent total electron spectra. The initial electron spectrum is a thermal spectrum peaks at $\gamma\sim 200$. $N(\gamma)$ is the particle number distribution. The dash line is the spectrum without cooling. Lower panel: time-dependent photon spectra from radio to near UV bands. Both panels are chosen at the labeled time steps in the unit of $\tau_{lc}$. \label{spectrum}}
\end{figure}

\begin{figure}[ht]
\centering
\includegraphics[width=0.5\linewidth]{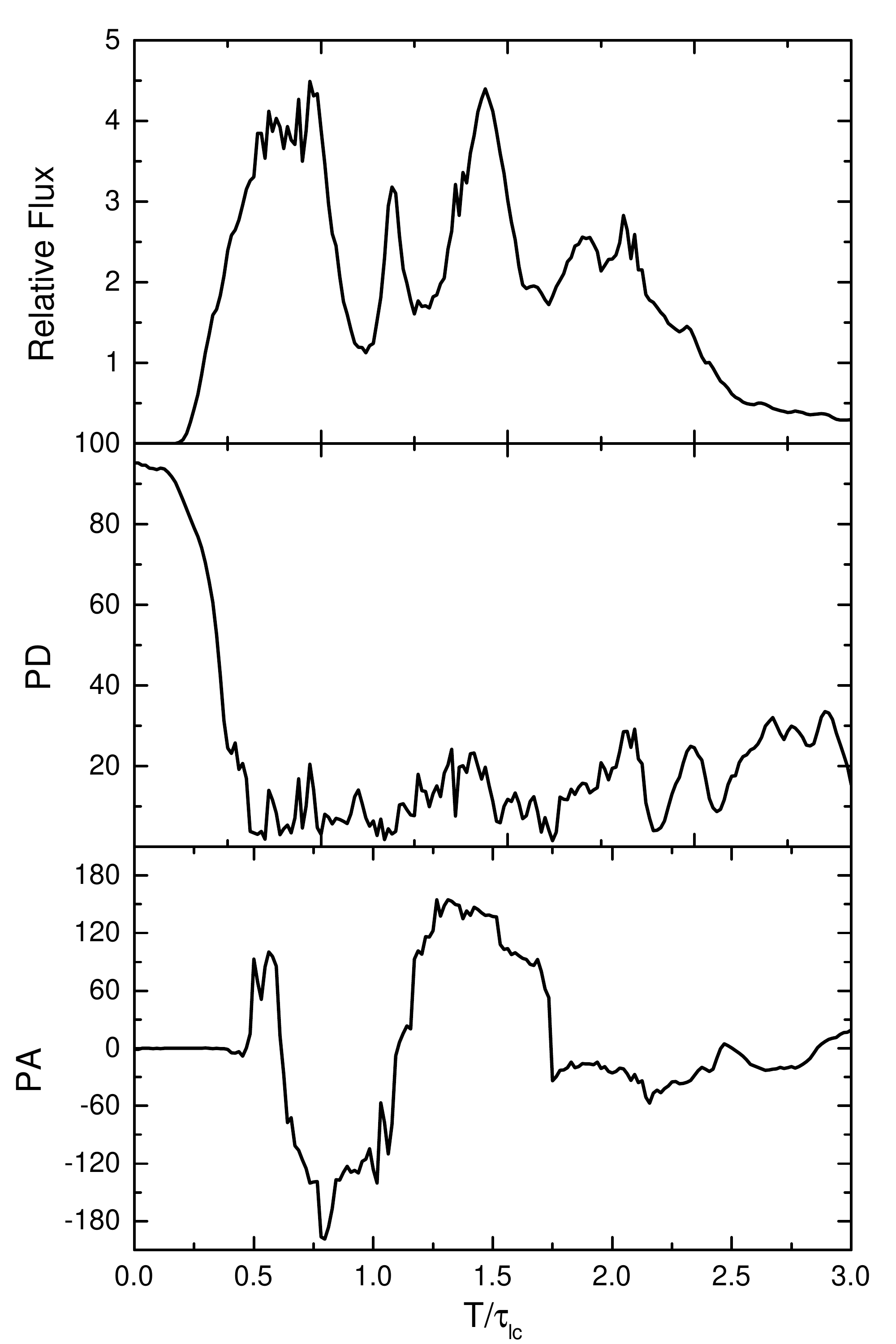}
\caption{Upper panel: relative flux in the optical band as a function of time. Middle panel: optical PD. Lower panel: optical PA. \label{lightcurve}}
\end{figure}

Blazar jets have a bulk Lorentz factor of a few tens and the plane of the reconnection layer in the blazar emission region can form different angles with the line of sight (LOS). Therefore, the viewing angle and Doppler boosting can affect the apparent magnetic field structure. For simplicity, here we assume that the reconnection layer forms in the $x-z$ plane and that we are observing reconnection site along the $y$ axis in its comoving frame, while the jet moves in the $z$ direction with a Lorentz factor of $10$. We Doppler boost all radiation signatures to the observer's frame with $\delta=10$. Fig. \ref{spectrum} shows time-dependent electron spectra and photon spectral energy distributions, and Fig. \ref{lightcurve} shows the light curves and polarization signatures in the optical band.

\begin{figure}[ht]
\centering
\includegraphics[width=1.0\linewidth]{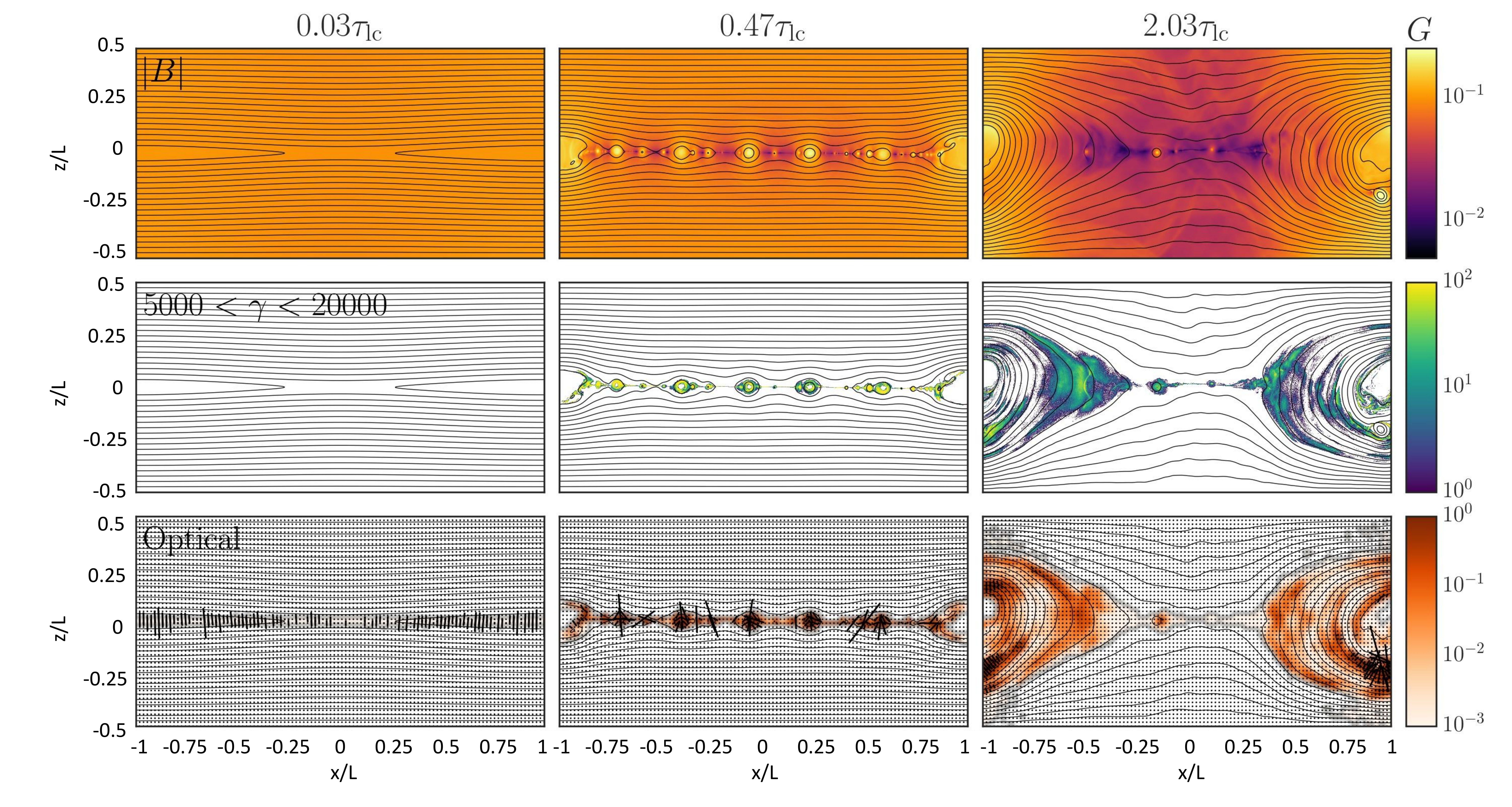}
\caption{Magnetic field strength (upper row), particle number density (middle row), and the polarized emission map (lower row) of the simulation region. In the lower row, the color indicates the total flux at each zone, while the segments represent the relative polarized flux (see Section 3.2 for its definition). Different columns represent snap shots of the simulation domain at different time steps. \label{pic}} 
\end{figure}

\begin{figure}[ht]
\centering
\includegraphics[width=1.0\linewidth]{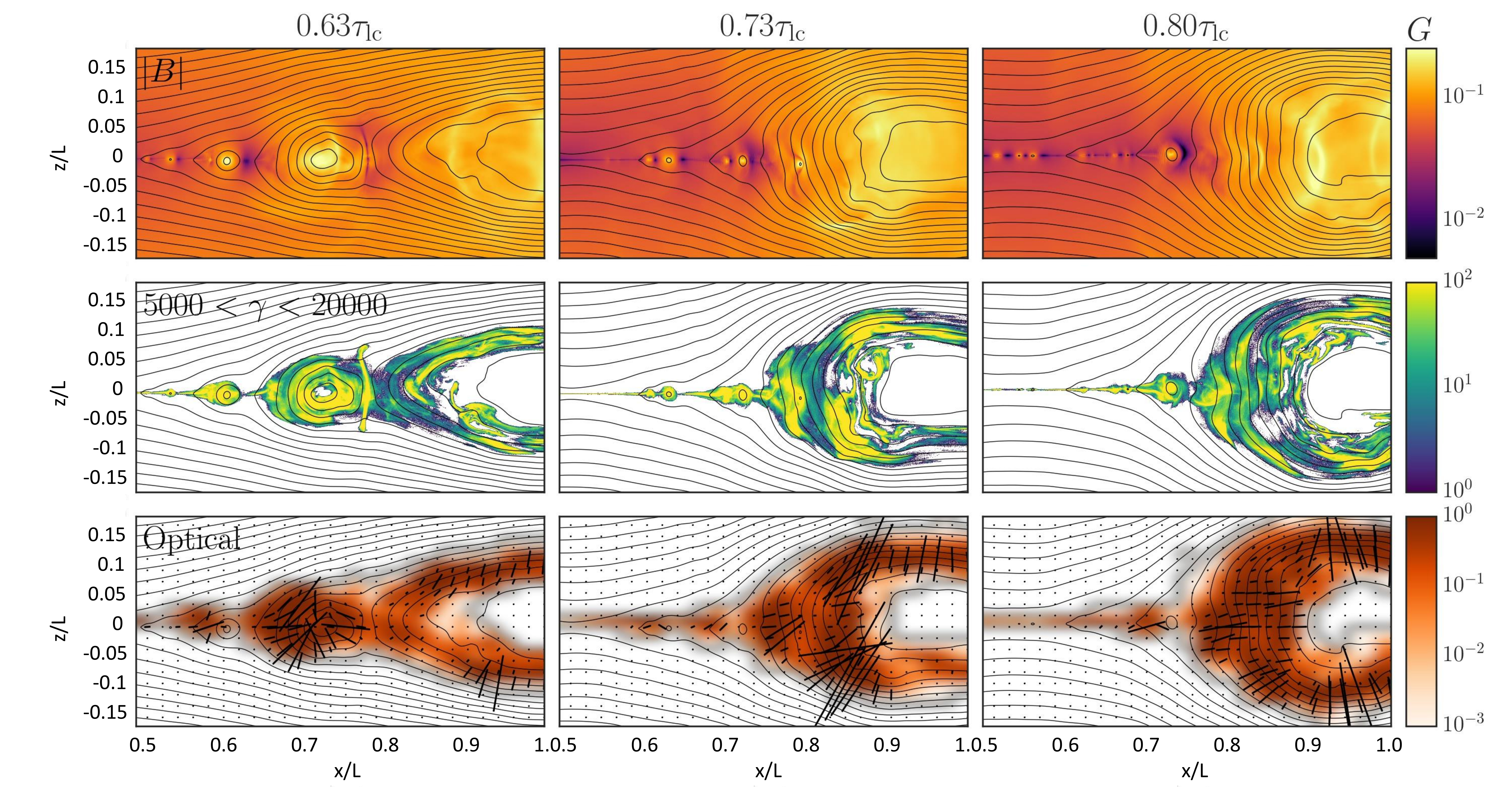}
\caption{Zoom-in view of one major plasmoid merger. Otherwise is the same as Fig. \ref{pic}. \label{resolve1}}
\end{figure}

\begin{figure}[ht]
\centering
\includegraphics[width=1.0\linewidth]{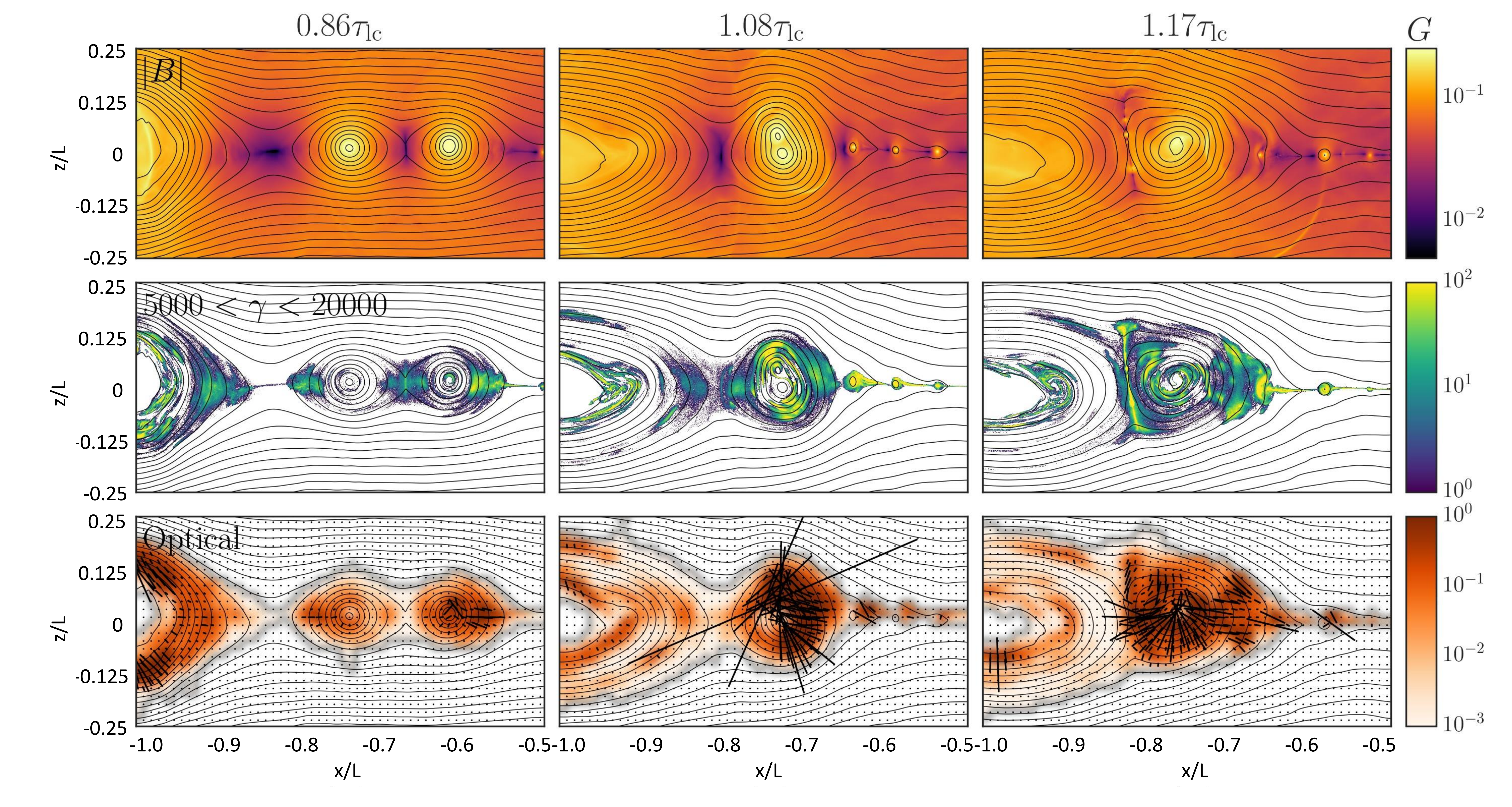}
\caption{Same as Fig. \ref{resolve1} for another major plasmoid merger. \label{resolve2}}
\end{figure}

\subsection{Spectra and Light Curves}

The onset of the magnetic reconnection is triggered by an initial perturbation. At $\sim 0.25\tau_{lc}$, magnetic field lines start to reconnect. The reconnection layer breaks into a series of fast moving plasmoids, and quickly accelerates nonthermal particles. These plasmoids are quasi-circular structures of nonthermal particles and magnetic field lines (Fig. \ref{pic} central column). The direction with which the magnetic field circles around the plasmoid is determined by the initial anti-parallel magnetic structure and is the same for all plasmoids along the reconnection layer. Due to their velocity differences, plasmoids can collide and merge into each other. During the plasmoid coalescence, since their magnetic fields are all clockwise, the merging site has an anti-parallel magnetic field component with a newly formed current sheet. This triggers secondary reconnection during the coalescence, leading to additional particle acceleration. This feature is clearly exhibited in Figs. \ref{resolve1} and \ref{resolve2}, which are the zoom-in figures of two plasmoid coalescences. Because of the periodic boundary in the $x$ direction, all plasmoids generated at the reconnection layer eventually merge to the big plasmoid at the $x$-axis boundary. After a significant amount of magnetic energy has been dissipated, the reconnection process saturates at $\sim 2.4\tau_{lc}$.

Magnetic reconnection quickly accelerates electron to a power-law distribution, while the radiative cooling cools high-energy electrons (Fig. \ref{spectrum}). The combined effect creates an overall broken power-law spectral shape. The spectral break marks the transition from slow cooling to fast cooling. Comparing to the same run without cooling (Fig. \ref{spectrum}), where the electron power-law turns over at $\sigma_e$, we find that the cooling limits the maximal electron energy. We estimate that the so-called synchrotron burnoff limit \citep[$\gamma_{rad}\sim 4.5\times 10^4$,][]{Uzdensky11}, where the Lorentz force equals to the synchrotron cooling, is comparable to $\sigma_e$. In our simulated optical light curve, on top of the first peak, we observe a series of sub-flares. This is because at the early stage of the reconnection, many small plasmoids quickly merge into each other, giving rise to a series of particle acceleration episodes at their secondary reconnection layers (Fig. \ref{pic} central column). Such phenomena are frequently observed during large blazar flares and the prompt phase of gamma-ray bursts. As the plasmoids become larger, the time intervals between mergers increase and the acceleration of particles in the reconnection layer cannot keep up with the radiative cooling, thus the optical flux drops. Later, when these larger plasmoids move closer and start to merge (Figs. \ref{resolve1} and \ref{resolve2}), island coalescence events produce a large amount of nonthermal electrons. This even dominates over the acceleration in the reconnection layer, resulting in several follow-up flares that can reach a flux level comparable to the first one. We notice that these flares are not result of the local Doppler boost of plasmoids (mini-jets), because we are observing in the $y$ direction, which is perpendicular to the plasmoid moving direction, and bulk Lorentz factors for our simulation setup and parameters are small, usually $\Gamma \lesssim 2$. Therefore, we conclude that the plasmoid mergers during magnetic reconnection events can lead to strongly variable light curves.

\subsection{Polarization Degree and Angle}

Magnetic reconnection can lead to strongly variable polarization signatures. In the last row of Figs. \ref{pic}, \ref{resolve1}, and \ref{resolve2}, we overlap the relative surface brightness with polarization vectors, whose direction represents the PA, and the length shows the local ``relative polarized flux''. The relative polarized flux is defined as the ratio of the local polarized flux to the global flux of the entire reconnection layer at the same time step. Thus, a longer polarization line at one cell in Fig. \ref{pic} means a larger polarization contribution to the total polarization signatures. Notice that the perpendicular polarization component in different cells can cancel each other if they arrive to the observer at the same time.

When there are no major plasmoid mergers, the relative polarized flux is distributed rather evenly along the layer and in all polarization directions. Given that the perpendicular polarization directions offset each other in the total polarization, the observed PD is very low, and the observed PA represents the chance residual. During the plasmoid mergers, however, the local relative polarized flux at the merger site dominates because of the additional particle acceleration. This results in a transient concentration of the polarized flux in similar polarization directions. Consequently, the PD increases temporarily. This feature is especially prominent in the late flares, which are triggered by relatively large plasmoid mergers. During these periods, the PD can rise up to $\sim 20\%$ for a considerably long time. However, when the merger moves toward completion, radiative cooling drives a sharp decline in the lightcurve and the PD returns to a low level. Therefore, we suggest that the PD is low and variable during magnetic reconnection.

When two plasmoids merge into each other, they can rotate with respect to each other (particularly clear in Fig. \ref{resolve2}). This results in flows of newly accelerated electrons moving along the reconnected magnetic field lines of the merging plasmoids. As the secondary reconnection layer is a highly dynamical region, the newly accelerated electrons are inhomogeneously distributed around this area. Therefore, we observe an asymmetry between the flow of nonthermal electrons that move clockwise and counterclockwise (Figs. \ref{resolve1} and \ref{resolve2} middle rows). The dominating flow however depends on the exact physical condition of the merging plasmoids, as we can see in Fig. \ref{resolve1}, the counterclockwise flow is stronger, while in Fig. \ref{resolve2}, the clockwise flow is stronger. Therefore, when two plasmoids merge, they can lead to a systematic rotation of the dominating PA in either direction. Since the PA swing originates from the asymmetric nonthermal electron bulk flow, the PA rotation amplitude of one plasmoid merger generally does not exceed $180^{\circ}$. However, there are a large number of plasmoid mergers in the reconnection layer. If successive dominating plasmoid mergers happen to have the same polarization rotating direction, the PA can continue to rotate in the same direction to make large-amplitude PA swings. Such situation may happen during the middle stage of the reconnection, when relatively large plasmoids merge into each other. One may expect when the dominating plasmoid merger succeeds a previous one, their PA may not be at the same position. Therefore, we see some bumps on the PA curve during large-amplitude PA rotation (Fig. \ref{lightcurve} lower panel).

\section{Discussion and Conclusion \label{discussion}}

Relativistic magnetic reconnection events may widely exist in magnetized plasmas in astrophysical systems, such as active galactic nuclei, gamma-ray bursts, and pulsar wind nebulae. Polarization signatures can explore the unique dynamical magnetic field evolution during reconnection events. Blazar jets are closely monitored by multi-wavelength campaigns with polarimetry, making them ideal targets to study reconnection physics. Simultaneous multi-wavelength light variability and polarization signatures shed light on the co-evolution of nonthermal particles and magnetic field in the blazar emission region. Recent blazar observations have revealed that optical PA swings are frequently associated with one or multiple multi-wavelength flares \citep{Marscher08,Marscher10,Blinov16,Blinov18}. Typically, the optical PD drops during the PA swing, and fluctuates at a low level \citep{Blinov16}. The amplitude of PA swings are usually around $\sim 180^{\circ}$ \citep{Blinov16}, but in rare cases much larger swings have been observed  \citep{Marscher10,Chandra15}. In addition, the PA can rotate in both directions in the same source and even during one blazar flare \citep{Chandra15}.

Theoretically, models of blazar PA swings generally fall into three categories. One scenario is the geometric effects, such as a bending jet \citep{Marscher08}, a moving blob along magnetic fields \citep{Marscher10}, or a rotating beam \citep{Lyutikov17}. These models can explain arbitrarily large PA swings associated with blazar flares. However, their PD has explicit patterns due to their respective geometric effects, contradicting to the observed low and fluctuating PD. Additionally, they favor that all PA swings in the same source generally rotate in the same direction. Another scenario is the stochastic processes in a turbulent magnetic field \citep{Marscher14}. This model features low and variable PD consistent with observations, but it rarely makes $\gtrsim 180^{\circ}$ PA swings. Furthermore, this model predicts very noisy PA swings, but observations have often seen rather smooth swings that disfavor a stochastic origin \citep{Kiehlmann17}. The third possibility is a regulated magnetic field alteration due to local energy dissipations, such as shocks or magnetic instabilities \citep{ZHC15,Nalewajko17}. Magnetohydrodynamic simulations have shown that shocks and kink instabilities can locally modify the magnetic field and dissipate jet energy, giving rise to correlated flares, low PD, and smoothly rotating PA. Nevertheless, the rotation amplitude is only $\sim 180^{\circ}$. Therefore, so far there is no satisfactory explanation for large-amplitude PA swings.

We employ a first-principle approach to study the radiation and polarization signatures of magnetic reconnection events, by combining PIC with radiative cooling and polarized radiation transfer simulations. During the magnetic reconnection, the reconnection layer breaks into a series of moving plasmoids containing accelerated nonthermal particles and quasi-circular magnetic field. This leads to the overall power-law spectral shape and low orderness of the magnetic field. Plasmoids can collide and merge into each other, forming secondary reconnection layers and efficiently accelerating particles. Therefore, the polarization at plasmoid merging sites dominates the observed signatures. During a dominating plasmoid merger, the newly accelerated electrons can flow as a bulk along the reconnecting quasi-circular magnetic field lines, lighting up the local polarization direction successively along the trajectory. This results in a systematic PA swing. A series of plasmoid mergers may then lead to large-amplitude PA swings in both directions. We emphasize the differences between the magnetic reconnection and turbulent magnetic field scenarios. In a turbulent magnetic field, polarization signatures are dominated by random walks of small polarization fluctuations. Thus we do not expect smooth and systematic patterns. In the magnetic reconnection scenario, while the plasmoid mergers can appear very stochastic and strongly dependent on the local plasma conditions, the polarization variations during one specific merger is quite systematic and can have a large PA shift. Therefore, the magnetic reconnection polarization signatures possess both the smooth patterns that are dominated by one merger and the bumpy patterns that occur when a new plasmoid merger succeeds the old one.

To summarize, our first-principle simulation based on integrated PIC and polarized radiation transfer suggest that the plasmoid coalescences during the reconnection can lead to multiple strong flares, low and fluctuating PD, as well as PA swings. These features are consistent with observations. We find that large-amplitude PA swings simultaneously with strong flares may be a unique signature of the relativistic magnetic reconnection in the blazar emission region.

\acknowledgments{We thank the anonymous referee for very helpful suggestions. HZ acknowledges support from Fermi Guest Investigator program Cycle 10, grant number 80NSSC17K0753. HZ and DG acknowledges support from NASA grants NNX16AB32G and NNX17AG21G. F. G. and X. L. acknowledge DOE OFES, and the support by the DOE through the LDRD program at LANL. Simulations are carried out on LANL clusters provided by LANL Institutional Computing.}

\clearpage


\begin{thebibliography}{}

\bibitem[Ackermann et al.(2016)]{Ackermann16} Ackermann, M., Anantua, R., Asano, K., et al.\ 2016, \apjl, 824, L20

\bibitem[Angelakis et al.(2016)]{Angelakis16} Angelakis, E., Hovatta, T., Blinov, D., et al.\ 2016, \mnras, 463, 3365 

\bibitem[Birn et~al. (2001)]{Birn01} Birn, J., Drake, J.~F., Shay, M.~A., et al. 2001, JGR, 106, 3715

\bibitem[Blandford \& Znajek(1977)]{Blandford77} Blandford, R.~D., \& Znajek, R.~L.\ 1977, \mnras, 179, 433 

\bibitem[Blinov et al.(2016)]{Blinov16} Blinov, D., Pavlidou, V., Papadakis, I.~E., et al.\ 2016, \mnras, 457, 2252 

\bibitem[Blinov et al.(2018)]{Blinov18} Blinov, D., Pavlidou, V., Papadakis, I., et al.\ 2018, \mnras, 474, 1296 

\bibitem[B{\"o}ttcher et al.(2013)]{Boettcher13} B{\"o}ttcher, M., Reimer, A., Sweeney, K., \& Prakash, A.\ 2013, \apj, 768, 54 

\bibitem[Bowers et~al. (2008)]{Bowers08} Bowers, K.~J., Albright, B.~J., Yin, L., Bergen, B., \& Kwan, T.~J.~T. 2008, PhPl, 15, 055703

\bibitem[Cerutti et al.(2012)]{Cerutti12} Cerutti, B., Werner, G.~R., Uzdensky, D.~A., \& Begelman, M.~C.\ 2012, \apjl, 754, L33 

\bibitem[Cerutti et al.(2013)]{Cerutti13} Cerutti, B., Werner, G.~R., Uzdensky, D.~A., \& Begelman, M.~C.\ 2013, \apj, 770, 147 

\bibitem[Chandra et al.(2015)]{Chandra15} Chandra, S., Zhang, H., Kushwaha, P., et al.\ 2015, \apj, 809, 130 

\bibitem[Giannios et al.(2009)]{Giannios09} Giannios, D., Uzdensky, D.~A., \& Begelman, M.~C.\ 2009, \mnras, 395, L29 

\bibitem[Gosling et al.(2005)]{Gosling05} Gosling, J.~T., Skoug, R.~M., McComas, D.~J., \& Smith, C.~W.\ 2005, Journal of Geophysical Research (Space Physics), 110, A01107 

\bibitem[Guo et al.(2014)]{Guo14} Guo, F., Li, H., Daughton, W., \& Liu, Y.-H.\ 2014, Physical Review Letters, 113, 155005

\bibitem[Guo et al.(2015)]{Guo15} Guo, F., Liu, Y.-H., Daughton, W., \& Li, H.\ 2015, \apj, 806, 167 

\bibitem[Guo et al.(2016)]{Guo16} Guo, F., Li, X., Li, H., et al.\ 2016, \apjl, 818, L9

\bibitem[Kiehlmann et al.(2017)]{Kiehlmann17} Kiehlmann, S., Blinov, D., Pearson, T.~J., \& Liodakis, I.\ 2017, \mnras, 472, 3589 

\bibitem[Landau \& Lifshitz (1975)]{Landau1975} Landau, L.~D., \& Lifshitz, E.~M. 1975, The classical theory of fields

\bibitem[Lyutikov \& Kravchenko(2017)]{Lyutikov17} Lyutikov, M., \& Kravchenko, E.~V.\ 2017, \mnras, 467, 3876 

\bibitem[Marscher et al.(2008)]{Marscher08} Marscher, A. P., et al., 2008, Nature, 452, 966

\bibitem[Marscher et al.(2010)]{Marscher10} Marscher, A.~P., Jorstad, S.~G., Larionov, V.~M., et al.\ 2010, \apjl, 710, L126 

\bibitem[Marscher(2014)]{Marscher14} Marscher, A. P., 2014, ApJ, 780, 87

\bibitem[Nalewajko(2017)]{Nalewajko17} Nalewajko, K.\ 2017, Galaxies, 5, 64 

\bibitem[Petropoulou et al.(2016)]{Petropoulou16} Petropoulou, M., Giannios, D., \& Sironi, L.\ 2016, \mnras, 462, 3325 

\bibitem[Phan et al.(2000)]{Phan00} Phan, T.~D., Kistler, L.~M., Klecker, B., et al.\ 2000, \nat, 404, 848 

\bibitem[Sironi \& Spitkovsky(2014)]{Sironi14} Sironi, L., \& Spitkovsky, A.\ 2014, \apjl, 783, L21 

\bibitem[Tian et al.(2014)]{Tian14} Tian, H., Li, G., Reeves, K.~K., et al.\ 2014, \apjl, 797, L14 

\bibitem[Uzdensky et al.(2011)]{Uzdensky11} Uzdensky, D.~A., Cerutti, B., \& Begelman, M.~C.\ 2011, \apjl, 737, L40 

\bibitem[Wang et al.(2016)]{Wang16} Wang, R., Lu, Q., Nakamura, R., et al.\ 2016, Nature Physics, 12, 263 

\bibitem[Werner et al.(2016)]{Werner16} Werner, G.~R., Uzdensky, D.~A., Cerutti, B., Nalewajko, K., \& Begelman, M.~C.\ 2016, \apjl, 816, L8 

\bibitem[Zhang et al.(2015)]{ZHC15} Zhang, H., Chen, X., B\"ottcher, M., Guo F., \& Li, H., 2015, ApJ, 804, 58

\bibitem[Zhang et al.(2016)]{ZHC16} Zhang, H., Deng, W., Li, H., \& B\"ottcher, M., 2016, ApJ, 817, 63

\bibitem[Zhang et al.(2017)]{ZHC17} Zhang, H., Li, H., Guo, F., \& Taylor, G.\ 2017, \apj, 835, 125 

\end{thebibliography}
\end{document}